\documentclass[12pt]{iopart}
\usepackage{amssymb}
\usepackage{amsfonts}
\usepackage{graphicx}
\usepackage{color}
\usepackage{cite}
\eqnobysec

\def\keywords#1{\vspace{10pt}
     \begin{indented}
     \item[]\rm Keywords: #1\par
     \end{indented}}

\def\be{\begin{equation}}
\def\ee{\end{equation}}
\def\bea{\begin{eqnarray}}
\def\eea{\end{eqnarray}}
\def\BD{\mathbf{\Delta}}

\def\CK{{\mathcal V}}
\def\CJ{{\mathcal J}}
\def\CK{{\mathcal K}}
\def\CT{{\mathcal T}}
\def\CS{{\mathcal S}}

\def\fT{{\mathfrak T}}
\def\fQ{{\mathfrak Q}}

\def\fS{{\mathfrak S}}
\def\fP{{\mathfrak P}}
\def\BD{\mathbf{\Delta}}
\def\scal{\stackrel{\rm s.l.}{=}}
\def\simscal{\stackrel{\rm s.l.}{\simeq}}

\begin{document}
\jl{1}

\title[Records for the number of distinct sites visited by a random walk]{Records for the 
number of distinct sites visited by a random walk on the fully-connected lattice}

\author{Lo\"\i c Turban}

\address{Groupe de Physique Statistique, D\'epartement P2M, Institut Jean Lamour,\\ Universit\'e de Lorraine, CNRS (UMR 7198), 
Vand\oe uvre l\`es Nancy Cedex, F-54506, France} 

\ead{loic.turban@univ-lorraine.fr}

\begin{abstract}
We consider a random walk on the fully-connected lattice with $N$ sites and study the time evolution 
of the number of distinct sites $s$ visited by the walker on a subset with $n$ sites. A record value 
$v$ is obtained for $s$ at a record time $t$ when the walker visits a site of the subset 
for the first time. The record time $t$ is a partial covering time when $v<n$ and a total 
covering time when $v=n$. The probability distributions for the number of records $s$, the record value $v$ 
and the record (covering) time $t$, involving $r$-Stirling numbers, are obtained using 
generating function techniques. The mean values, variances and skewnesses are deduced 
from the generating functions. In the scaling limit the probability distributions for 
$s$ and $v$ lead to the same Gaussian density. The fluctuations of the record time $t$ 
are also Gaussian at partial covering, when $n-v={\mathrm O}(n)$. They are distributed according to the type-I Gumbel extreme-value distribution at total covering, when $v=n$. A discrete sequence of generalized Gumbel distributions, indexed by $n-v$, is obtained at almost total covering, when $n-v={\mathrm O}(1)$. These generalized Gumbel distributions are crossing over to the Gaussian distribution when $n-v$ increases. 
\end{abstract}




\section{Introduction} 
\begin{figure}[!t]
\begin{center}
\includegraphics[width=6cm,angle=0]{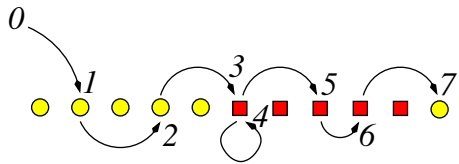}
\end{center}
\vglue -.0cm
\caption{In this example the fully-connected lattice has $N=11$ sites among which $n=5$ (squares) belong to the subset.
At $t=0$ the walker is outside the lattice. A first record occurs at the third step with record time $t=3$ and 
record value $v=1$. There is no record at the next step, $t=4$, since the walker visits this site for 
the second time. The two following steps, corresponding to the visit of new sites belonging to the 
subset, bring two new records with, respectively,  record time $t=5$ and value $v=2$,  record time 
$t=6$ and value $v=3$. At $t=7$, which is not a record time, the number of records is $s=3$. 
The evolution of $s$ as a function of $t$ is shown in figure~\ref{fig-2}(a).
\label{fig-1}  
}
\end{figure}

In a recent work \cite{turban14} (referred to as I) exact results for the statistics of the number 
of distinct sites $s$ visited by a random walker up to time $t$ on the fully-connected lattice with 
$N$ sites were presented. Discrete probability distributions, involving Stirling numbers of the 
second kind, were obtained leading to Gaussian distributions in the continuum, scaling limit. 
The present work is a continuation of I in which we study the statistical properties of the records 
associated with these numbers $s$. The walk is still taking place on the fully-connected lattice with $N$ 
sites but $s$ is now the number of distinct sites visited by the walker, belonging to a subset with 
$n$ sites (see~\cite{weiss82} for a study of this problem in dimension $d=1$ to~3). A record is 
established each time the walker visits a new site among the $n$ sites of the subset 
(see figures~\ref{fig-1} and~\ref{fig-2}(a)). To each record we associate a record time $t$ 
and a record value $v$ whereas $s$ gives the number of records. We study the discrete probability 
distributions associated with the stochastic variables $s$, $v$ and $t$, and their scaling behaviour.

The study of records~\cite{chandler52,glick78,arnold98,nevzorov01,nagaraja03} has been a subject of 
renewed interest in the physics community during the last  
years~\cite{schmittmann99,krug07,franke10,wergen11,franke12,majumdar12,wergen13,godreche14,bennaim14,fortin15}. 
In the visited sites problem the records are not standard ones since $s$ is a non-decreasing random variable. 
A record time corresponding to the total covering of a graph with $N$ vertices by a random walk is 
usually called the covering (or cover) time of the graph. In the following we use indifferently 
the terms `record time' or `covering time'. We shall have to distinguish between the {\em partial covering time} 
when the record value $v$ is such that $\lim_{v,n\to\infty}v/n=x<1$ (i.e., $n-v={\mathrm O}(n)$), the {\em total 
covering time} when $x=1$ with $v=n$ and the {\em almost total covering time} when $x=1$ with $n-v={\mathrm O}(1)$.

Exact results for the mean covering time have been obtained on a lattice with $N$ sites in one dimension 
($1D$) with either periodic or reflecting boundary conditions~\cite{yokoi90}:
\bea
&\overline{t_N^{(P)}}=\frac{1}{2}N(N-1)\qquad  ({\rm periodic})\,,\nonumber\\
&\overline{t_N^{(R)}(i)}=N(N-1)+(i-1)(N-i)\qquad ({\rm reflecting})\,.
\label{tNpr}
\eea
In the reflecting case $i$ gives the initial position of the walker. The typical size 
of the domain explored by the random walker at time $t$ grows as $\sqrt{t}$ which explains 
the long time quadratic growth with $N$. The same problems have been recently solved 
in the case of a persistent random walk~\cite{chupeau14}. Exact results have also been 
obtained in $1D$ for the mean partial covering time and the mean random covering 
time~\cite{nascimento01} which is the time needed to visit a given fraction of the $N$ sites chosen at random.

On the basis of Monte Carlo simulations, the following expressions were 
conjectured~\cite{nemirovsky90} for the mean covering time for $N\gg1$ in higher dimensions:
\bea
\overline{t_N}&=A_2N\ln^2N[1+{\mathrm O}(1/\ln N)]\qquad (d=2)\,,\nonumber\\
\overline{t_N}&=A_dN\ln N[1+{\mathrm O}(1/\ln N)]\qquad (d\geq3)\,.
\label{tNd}
\eea
Logarithmic corrections can be traced to multiple visits of the same sites and this effect is stronger in $2D$. 

Actually the leading contribution for $d\geq3$ has been derived earlier using methods of probability 
theory~\cite{aldous83}. The amplitude $A_d$ can be expressed in terms of the probability of return 
to the origin. The conjecture \eref{tNd} was also indirectly confirmed analytically for 
$d\geq2$~\cite{brummelhuis91,brummelhuis92} with the following values for the amplitudes on 
hypercubic lattices with periodic boundary conditions:
\be
A_d=\left\{
\begin{array}{ll}
1/\pi=0.318\ldots\,, & d=2\,,\\
1.516\ldots\,,       & d=3\,,\\
1.239\ldots\,,       & d=4\,.
\end{array}
\right.
\label{Ad}
\ee
In mean-field, or $d=\infty$, the amplitude is simply given by
$A_\infty=1$~\cite{aldous83,brummelhuis91,brummelhuis92,nemirovsky91,lovasz96}.

The mean covering time for $k$ visits has been studied in $1D$ through Monte Carlo 
simulations~\cite{mirasso91}. The same method has been 
used to evaluate partial and random covering times in $2D$~\cite{coutinho94}.

Some exact results are known for the probability distribution of the total covering time. 
An exact closed-form analytical expressions has been derived for a random walk on an 
arbitrary graph~\cite{zlatanov09}. 
The expression for the complete graph (or fully-connected lattice) is in agreement 
with a conjecture based on small-size exact enumerations~\cite{nemirovsky91}. 
There it was noticed that the probability to have a total covering time equal to its minimum value is 
given by the ratio of the number of Hamiltonian walks $Z_{\rm HW}$ to the number of random walks  
$Z_{\rm RW}$ with $N$ steps. Then using a $1/d$ expansion of $Z_{\rm HW}$
for the hypercubic lattice~\cite{nemirovsky88,nemirovsky89} it was shown that the first non-vanishing 
correction to $Z_{\rm HW}/Z_{\rm RW}$ is of order $1/d^2$.

It has been rigorously shown recently~\cite{belius13} that the fluctuations of the total 
covering time for a random walk on the discrete torus in dimension $d\geq3$ are governed 
by the Gumbel distribution~\cite{gumbel35,gumbel04} in the scaling limit. This result 
was earlier conjectured in~\cite{aldous02}.

Our main results for random walks on the fully-connected lattice can be summarized as follows. 
The number of records $s$ established up to time $t$ is distributed according 
to~\footnote[1]{Note that $s$ is also the number of distinct sites visited on 
the subset up to time $t$. The distribution~\eref{s} was obtained in I for $n=N$}   
\be
S_{N,n}(s,t)=\frac{n^{\underline{s}}}{N^t}{N-n+t\brace N-n+s}_{N-n}\,.
\label{s}
\ee
In this expression $N$ is the lattice size, $n$ the subset size, $n^{\underline{s}}$ is a 
falling factorial power~(\cite{graham94}, p 47) and ${k\brace m}_r$ an $r$-Stirling number 
of the second kind~\cite{broder84}.

The record value $v$ for a given record time $t$ is distributed according to
\be
V_{N,n}(v,t)=\frac{(n-1)^{\underline{v-1}}}{(N\!-1)^{t-1}}{N\!-n+t-1\brace N\!-n+v-1}_{N-n}\,,
\label{v}
\ee
whereas the record time $t$ for a given record value $v\leq n$ is distributed according to 
\be
T_{N,n}(v,t)=\frac{n^{\underline{v}}}{N^t}{N-n+t-1\brace N-n+v-1}_{N-n}\,.
\label{t1}
\ee 

In the scaling limit, indicated by `\,s.l.' ($N\to\infty$, $n\to\infty$, $t\to\infty$ or $v\to\infty$,  with fixed ratios
$n/N\scal f$, $t/N\scal w$ or $v/n\scal x$), $s$ and $v$ have the same mean values $\overline{s_{N,n}(t)}$ 
and $\overline{v_{N,n}(t)}$ such that
\be
\frac{\overline{s_{N,n}(t)}}{n}\scal\frac{\overline{v_{N,n}(t)}}{n}\scal 1-\e^{-w}\,.
\label{sv1}
\ee
The probability distributions $S_{N,n}(s,t)$ and $V_{N,n}(v,t)$ lead to the same centered Gaussian density in the reduced variable
\be
\frac{s-\overline{s_{N,n}(t)}}{n^{1/2}}\scal\frac{v-\overline{v_{N,n}(t)}}{n^{1/2}}\scal\sigma\,.
\label{sv2}
\ee
The mean partial covering time $\overline{t_{N,n}(v)}$ behaves as:
\be
\frac{\overline{t_{N,n}(v)}}{N}\scal-\ln(1-x)\qquad (x<1)\,.
\label{t2}
\ee
The probability distribution $T_{N,n}(v,t)$ leads to a centered Gaussian density in the reduced variable
\be
\frac{t-\overline{t_{N,n}(v)}}{N^{1/2}}\scal\tau\qquad (x<1)\,,
\label{t3}
\ee
at partial covering.
The scaling behaviour is different at total covering. The mean value of the covering time is given by
\be
\overline{t_{N,n}(v)}\simeq N(\ln n+\gamma)\qquad (v=n\gg1)\,,
\label{t4}
\ee
where $\gamma=0.577\ 215\ 665\ldots$ is the Euler--Mascheroni constant.
The scaling limit of $NT_{N,n}(v,t)$ is the type-I Gumbel extreme value distribution in the reduced variable
\be
\frac{t-\overline{t_{N,n}(v)}}{N}\scal\tau'\,.
\label{t5}
\ee
The crossover from Gumbel to Gauss, in the vicinity of total covering when $u=n-v={\mathrm O}(1)$, 
occurs via a discrete sequence of generalized Gumbel distributions~\cite{ojo01}, indexed by the deviation 
$u$ from  total covering.

The outline of the paper is as follows. In section~2 the discrete probability distributions 
for the record numbers, record values and record times are obtained using generating functions 
techniques. Their moments are calculated in section~3, leading to the mean values, variances 
and skewnesses. The scaling limit is studied in section~4. The results are discussed
in section~5. Detailed calculations are presented in five appendices.

\section{Discrete probability distributions associated with records}
We consider a random walk on the fully-connected lattice with $N$ sites (see figure~1 in~I). 
The random steps are towards anyone of the $N$ sites, with probability $1/N$. The walker is 
outside the lattice at $t=0$ and the first random step on the lattice takes place at $t=1$. 
We study the properties of the number $s$ of distinct sites visited up to time $t$ on a fixed 
subset of $n$ sites arbitrarily chosen among the $N$. Since $s$ is a non-decreasing function 
of time, a new record for $s$ is established each time a new site is visited on the subset 
of $n$ sites.  Thus the value of $s$ at time $t$ also gives the number of records established 
up to time $t$. A record is characterized by its value $v$ and its time $t$, i.e., $v$ 
is the record value taken by $s$ at the record time $t$ (see figures~\ref{fig-1} and~\ref{fig-2}(a)).

\subsection{Generating functions}
In this section we are looking for the generating functions for the probability distributions 
$T_{N,n}(v,t)$ and $S_{N,n}(s,t)$. The first one gives the probability to establish a new record 
with value $v$ at time $t$ while the second gives the probability to have visited $s$ distinct 
sites (established $s$ records) up to time $t$. 

\begin{figure}[!t]
\begin{center}
\includegraphics[width=11cm,angle=0]{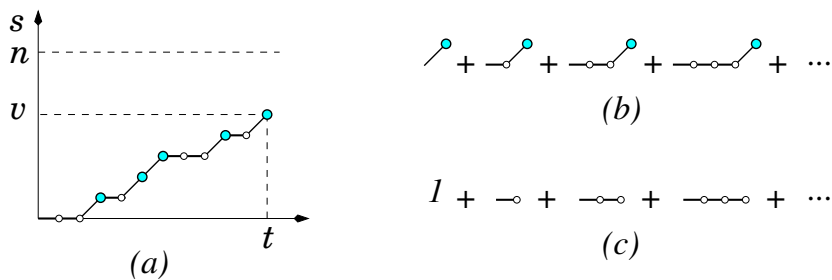}
\end{center}
\vglue -.0cm
\caption{
\label{fig-2} A time evolution of the number of distinct sites visited is sketched in (a). A small circle 
corresponds either to a step outside the subset of $n$ sites or on a previously visited site inside 
the subset. Bigger circles correspond to a record when the walker visits a new site inside the subset 
of $n$ sites. The sequence shown ends with a record with value $v$ at time $t$. The diagrams in (b) 
and (c) give the contributions to the generating function of the probability distribution 
$J_{N,n}(k,t)$ for the record lifetimes in~\protect\eref{CJz-1} and~\protect\eref{CKz}, respectively. 
}
\end{figure}

Let us first write the ordinary generating function for the probability distribution $J_{N,n}(s,l)$ 
of the lifetime $l$ when the record number is $s$: 
\be
\CJ_{N,n}(s,z)=\sum_{l=1}^\infty J_{N,n}(s,l)\,z^l=z\left(\frac{n-s}{N}\right)\CK_{N,n}(s,z)\,.
\label{CJz-1}
\ee
In this expression, the first two factors give the contribution of the record-breaking final step 
(bigger circles) in the diagrams of figure~\ref{fig-2}(b), occurring with probability $(n-s)/N$. 
The last one is the generating function which corresponds to the diagrams of figure~\ref{fig-2}(c)
\be
\CK_{N,n}(s,z)=\sum_{l=0}^\infty\left(\frac{N-n+s}{N}\right)^l z^l=\frac{N}{N-z(N-n+s)}\,,
\label{CKz}
\ee
and sums the contributions of sequences of $l$ steps, each with weight $z$ (smaller circles), 
for which the number of visited sites in the subset remains equal to $s$, with probability 
$(N-n+s)/N$. Inserting~\eref{CKz} into~\eref{CJz-1}, one obtains:
\be
\CJ_{N,n}(s,z)=\frac{z(n-s)}{N-z(N-n+s)}\,.
\label{CJz-2}
\ee

Any history in the $(s,t)$-plane, ending with a record at $(v,t)$ as shown in figure~\ref{fig-2}(a), 
can be decomposed into subsequences of steps at level $s$ ending with a record at $s+1$ as shown in 
figure~\ref{fig-2}(b), with $s$ increasing from $0$ to $v-1$. The record time $t$ is the sum of the 
lifetimes $l$ of the previous records. These are distributed according to the different distributions 
$J_{N,n}(s,l)$ for $s=0$ to $v-1$. {\em Thus the record times $t$ are the sum of $v$ 
independent, differently distributed random variables}. 

All the possible histories are taken into account if the generating function for $T_{N,n}(v,t)$ is 
written as the product of the generating functions for the lifetimes of the successive records, 
$\CJ_{N,n}(s,z)$, for $s=0$ to $v-1$:
\be\fl
\CT_{N,n}(v,z)=\sum_{t=1}^\infty T_{N,n}(v,t)\,z^t=\prod_{s=0}^{v-1}\CJ_{N,n}(s,z)
=\frac{n^{\underline{v}}\,(z/N)^v}{\prod_{s=0}^{v-1}\left[1-(z/N)(N-n+s)\right]}\,.
\label{CTz}
\ee
In the final expression $n^{\underline{v}}=n(n-1)\cdots(n-v+1)$ is a falling factorial power~\cite{graham94}.

For $S_{N,n}(s,t)$ the sequence of steps may or may not end on a record. This is taken into account 
in the generating function through a multiplication by $\CK_{N,n}(v,z)$ giving:
\be\fl
\CS_{N,n}(s,z)\!=\!\sum_{t=0}^\infty S_{N,n}(s,t)\,z^t\!=\!\CT_{N,n}(s,z)\,\CK_{N,n}(s,z)
\!=\!\frac{n^{\underline{s}}\,(z/N)^s}{\prod_{s'=0}^{s}\left[1\!-\!(z/N)(N-n+s')\right]}\,.
\label{CSz}
\ee

\subsection{Probability distributions}
The probability distribution  of the lifetime $l$ of record number $s$ is the coefficient of $z^l$ 
in the expansion of $\CJ_{N,n}(s,z)$ given by~\eref{CJz-2}:
\be
J_{N,n}(s,l)=\left(1-\frac{n-s}{N}\right)^{l-1}\left(\frac{n-s}{N}\right)\,.  
\label{Jsl}
\ee
This is the geometric distribution with parameter $p=\frac{n-s}{N}$.

The expressions obtained for $\CT_{N,n}(v,z)$ and $\CS_{N,n}(s,z)$ are closely related to the
generating function of the $r$-Stirling numbers of the second kind (see~\cite{broder84}, equation~(25))
\be
\sum_{l=0}^\infty{r+l\brace r+m}_r z^l=\frac{z^m}{\prod_{k=0}^{m}\left[1-z(r+k)\right]}\,,
\qquad r\geq 0,\  m\geq 0\,.
\label{rStir-1}
\ee
These numbers can be expressed in terms of ordinary Stirling numbers of the second 
kind ${k\brace m}$~\cite{graham94} through (\cite{broder84}, equation~(32))
\be
{r+l\brace r+m}_r=\sum_{k=0}^l{l\choose k}{k\brace m}\, r^{l-k}\,,
\label{rStir-2}
\ee
so that
\be
{r+l\brace r+m}_{r=0}={l\brace m}\,.
\label{rStir-3}
\ee
Comparing equations~\eref{CTz} and~\eref{CSz} to~\eref{rStir-1}, one obtains the joint probability distribution
\be
T_{N,n}(v,t)=\frac{n^{\underline{v}}}{N^t}{N-n+t-1\brace N-n+v-1}_{N-n}
\label{Tvt-1}
\ee
for the record value $v$ and the record time $t$ and the probability distribution
\be
S_{N,n}(s,t)=\frac{n^{\underline{s}}}{N^t}{N-n+t\brace N-n+s}_{N-n}
\label{Sst}
\ee
for the number of distinct sites $s$ visited up to time $t$ on the $n$-site subset of 
the fully-connected lattice with a total of $N$ sites.
When $n=N$ in~\eref{Sst} equation~(2.8) of I is recovered.

The last probability distribution satisfies the master equation
\be
S_{N,n}(s,t)=\frac{N-n+s}{N}S_{N,n}(s,t-1)+\frac{n-s+1}{N}S_{N,n}(s-1,t-1)\,,
\label{masteq-1}
\ee
for $t>0$, with $S_{N,n}(s,0)=\delta_{s,0}$ and $S_{N,n}(s<0,t)=0$.
The first term on the right, for which $s$ remains unchanged, corresponds to steps outside 
the subset of $n$ sites or on one of the $s$ sites already visited in the subset. 
With the second term the number of visited sites increases from $s-1$ to $s$ for steps 
on one of the $n-s+1$ sites not yet visited in the subset of $n$ sites at time $t-1$. 

Let us now look at the properties of $T_{N,n}(v,t)$. The probability distribution vanishes 
when $v>n$ due to the falling factorial and when $v>t$ due to the $r$-Stirling number. 
When summed over $t$, it gives the probability to establish a record with value $v$ so that:
\be
\sum_{t=1}^\infty T_{N,n}(v,t)=\left.\CT_{N,n}(v,z)\right|_{z=1}=
\left\{ 
\begin{array}{ll}
1 & \mbox{if $v\leq n$}\\
0 & \mbox{otherwise}
\end{array}
\right.
\label{norm}
\ee
It follows that $T_{N,n}(v,t)$ is the properly normalized probability distribution for 
the records times $t$ at a given value of $v\leq n$.
When summed over $v$, it gives the probability to establish a new record at time $t>0$:
\be
 B_{N,n}(t)=\sum_{v=1}^n T_{N,n}(v,t)\,.
\label{Bt-1}
\ee
Thus the normalized probability distribution for the value $v$ of a record established at time $t$ reads:
\be
V_{N,n}(v,t)=\frac{T_{N,n}(v,t)}{\sum_{v=1}^n T_{N,n}(v,t)}=\frac{T_{N,n}(v,t)}{B_{N,n}(t)}\,.
\label{Vvt-1}
\ee

To close this section let us look for the expression of the joint probability distribution in the simple case where $v=1$.
Then~\eref{Tvt-1} leads to
\be
T_{N,n}(1,t)=\frac{n}{N^t}{N-n+t-1\brace N-n}_{N-n}\,.
\label{T1t-1}
\ee
Making use of~\eref{rStir-2} with ${k\brace 0}=\delta_{k,0}$ gives ${r+l\brace r}_r=r^l$ and one obtains:
\be
T_{N,n}(1,t)=\left(1-\frac{n}{N}\right)^{t-1}\frac{n}{N}=J_{N,n}(0,t)\,.
\label{T1t-2}
\ee
This is the probability to have the first $t-1$ steps outside the subset followed by a visit to one of the $n$ sites of 
the subset at the record time $t$.

\section{Moments of the probability distributions}

\subsection{Moments of $J_{N,n}(s,l)$}
Using the expression of the geometric distribution in~\eref{Jsl} a direct calculation leads to the first two moments:
\be\fl
\overline{l_{N,n}(s)}\!=\!\sum_{l=1}^\infty l J_{N,n}(s,l)\!=\!\frac{N}{n\!-\!s}\,,\qquad 
\overline{l^2_{N,n}(s)}\!=\!\sum_{l=1}^\infty l^2 J_{N,n}(s,l)\!=2\!\left(\frac{N}{n-s}\right)^2\!\!-\!\frac{N}{n\!-\!s}\,.
\label{lNns}
\ee
The mean value increases with $s$ and is equal to $N$ when $s=n-1$, i.e., for the last lifetime 
before total covering of the subset. The variance is given by:
\be
\overline{\Delta l^2_{N,n}(s)}=\overline{l^2_{N,n}(s)}-\overline{l_{N,n}(s)}^2
=\left(\frac{N}{n-s}\right)^2-\frac{N}{n-s}\,.
\label{dlNn2s}
\ee
It is equal to $N(N-1)$ when $s=n-1$.

\subsection{Moments of $S_{N,n}(s,t)$}
Let us start with the number of distinct sites $s$ visited (number of records) 
up to time $t\geq0$. The moments of $S_{N,n}(s,t)$ can be deduced from the bivariate 
generating function obtained in appendix A:
\be
\CS_{N,n}(y,z)=\sum_{s=0}^\infty y^s\sum_{t=0}^\infty \frac{z^t}{t!}S_{N,n}(s,t)=\e^{z(1-n/N)}\left[1+y(\e^{z/N}-1)\right]^n\,.
\label{CSyz-4}
\ee
The mean number of distinct sites visited on the subset of $n$ sites up to time $t$ is 
given by the coefficient of $z^t/t!$ in the $y$-derivative of $\CS_{N,n}(y,z)$ at $y=1$:
\be
\overline{s_{N,n}(t)}=\left[\frac{z^t}{t!}\right]\left.\frac{\partial
\CS_{N,n}}{\partial y}\right|_{y=1}=n\left[1-\left(\frac{N-1}{N}\right)^t\right]\,.
\label{sNnt}
\ee
It is the product of the number of sites in the subset by the probability that a given site 
has been visited up to time $t$, $1-(1-1/N)^t$. 
The second moment reads:
\bea
\overline{s^2_{N,n}(t)}&=\left[\frac{z^t}{t!}\right]\left.\frac{\partial}{\partial y}
\left[y\,\frac{\partial\CS_{N,n}}{\partial y}\right]\right|_{y=1}\nonumber\\
&=n^2-n(2n-1)\left(\frac{N-1}{N}\right)^t+n(n-1)\left(\frac{N-2}{N}\right)^t\,.
\label{sNn2t}
\eea
Combining these results, the following expression is obtained for the variance:
\be\fl
\overline{\Delta s^2_{N,n}(t)}=n\left(\frac{N-1}{N}\right)^t\!\!+n(n-1)
\left(\frac{N-2}{N}\right)^t\!\!-n^2\left(\frac{N-1}{N}\right)^{2t}\!\!\!\!.
\label{dsNn2t}
\ee
Note that the variance is not proportional to $n$, which indicates a 
correlation effect. The visits of the different sites of the subset are not independent:
when one site is visited in a given step, the others are evidently not.

The third moment is given by:
\bea
\fl\overline{s^3_{N,n}(t)}&=\left[\frac{z^t}{t!}\right]\left.\frac{\partial}{\partial y}
\left[y\,\frac{\partial}{\partial y}\left(y\,\frac{\partial\CS_{N,n}}{\partial y}
\right)\right]\right|_{y=1}=n^3-\left[3n^2(n-1)+n\right]\left(\frac{N-1}{N}\right)^t\nonumber\\
\fl&\ \ \ \ \ \ +3n(n-1)^2\left(\frac{N-2}{N}\right)^t-n(n-1)(n-2)\left(\frac{N-3}{N}\right)^t\,.
\label{sNn3t}
\eea
It is needed to evaluate the skewness $\gamma_1$ measuring the asymmetry of the distribution. 
For a random variable $X$ the skewness is the third centered moment, normalized by the 
standard deviation to the third power~\cite{abramowitz72}:
\be
\gamma_1=\frac{\overline{(X-\overline{X})^3}}{(\overline{\Delta X^2})^{3/2}}
=\frac{\overline{X^3}-3\overline{X}\,\overline{\Delta X^2}-\overline{X}^3}{(\overline{\Delta X^2})^{3/2}}\,.
\label{gamX}
\ee
The skewness of $S_{N,n}(v,t)$ follows from previous results in equations~\eref{sNnt},~\eref{dsNn2t} and~\eref{sNn3t}:
\bea
\fl\gamma^{(S)}_{1\,N,n}(t)&=\frac{1}{n^{1/2}}\,\frac{-p_1+3np_1^2-3(n-1)p_2
-(n-1)(n-2)p_3+3n(n-1)p_1p_2-2n^2p_1^3}
{\left[\,p_1+(n-1)p_2-np_1^2\,\right]^{3/2}}\nonumber\\
\fl &\ \ \ \ \mbox{where}\ p_k=\left(\frac{N-k}{N}\right)^t \,.
\label{gamS}
\eea

\subsection{Moments of $V_{N,n}(v,t)$}
In order to study the moments of $V_{N,n}(v,t)$ let us first look closer at the relations 
between $T_{N,n}$, $V_{N,n}$ and $S_{N,n}$.
First, the probability to establish a new record with record value $v$ and record time $t$ 
is also the probability to have visited $v-1$ distinct sites at time $t-1$ and 
to visit a new site among the $n-v+1$ remaining ones at time $t$, thus we have
\be\fl
T_{N,n}(v,t)=S_{N,n}(v-1,t-1)\,\frac{n-v+1}{N}=S_{N,n}(v,t)-\frac{N-n+v}{N}\,S_{N,n}(v,t-1)\,,
\label{Tvt-2}
\ee
where the master equation~\eref{masteq-1} has been used.
Inserting the last expression in~\eref{Bt-1} leads to
\bea
\fl B_{N,n}(t)&=\sum_{v=1}^nS_{N,n}(v,t)-\frac{N-n}{N}\sum_{v=1}^nS_{N,n}(v,t-1)
-\frac{1}{N}\sum_{v=1}^nv\,S_{N,n}(v,t-1)\nonumber\\
\fl&=\underbrace{\sum_{v=0}^nS_{N,n}(v,t)}_1-\frac{N-n}{N}\underbrace{\sum_{v=0}^n
S_{N,n}(v,t-1)}_1-\frac{1}{N}\sum_{v=1}^nv\,S_{N,n}(v,t-1)\nonumber\\
\fl&=\frac{n}{N}-\frac{\overline{s_{N,n}(t-1)}}{N}=\frac{n}{N}\left(\frac{N-1}{N}\right)^{t-1}\,,
\label{Bt-2}
\eea
when $t>0$. The contribution of the terms $v=0$ added in the two first sums of the second line vanishes 
since $S_{N,n}(0,t)=\frac{N-n}{N}S_{N,n}(0,t-1)$. The final expression, which follows from~\eref{sNnt}, 
is the product of the probability to visit a site belonging to the subset at $t$, by the probability 
that it has never been visited before.
Making use of this expression in~\eref{Vvt-1} finally gives:
\bea
V_{N,n}(v,t)&=\frac{N}{n}\left(\frac{N}{N-1}\right)^{t-1}\frac{n^{\underline{v}}}{N^t}{N-n+t-1\brace N-n+v-1}_{N-n}\nonumber\\
&=\frac{(n-1)^{\underline{v-1}}}{(N\!-1)^{t-1}}{N\!-n+t-1\brace N\!-n+v-1}_{N-n}\!\!\!\!\!\!=S_{N\!-1,n-1}(v\!-\!1,t\!-\!1)\,.
\label{Vvt-2}
\eea
Since $S_{N,n}(s,0)=\delta_{s,0}$, the last equation gives $V_{N,n}(v,1)=\delta_{v,1}$, as expected.

According to~\eref{Vvt-2}, $V_{N,n}(v,t)$ is given by $S_{N-1,n-1}(s,t-1)$ at $s=v-1$. Thus their mean values are shifted:
\be
\overline{v_{N,n}(t)}=\overline{s_{N-1,n-1}(t-1)}+1\,.
\label{vNnt}
\ee
The variance and the skewness, which are both functions of $v-\overline{v_{N,n}(t)}$ are not 
affected by the shift. They are simply given by:
\be\fl
\overline{\Delta v^2_{N,n}(t)}=\overline{\Delta s^2_{N-1,n-1}(t-1)}\,,\qquad\gamma^{(V)}_{1\,N,n}(t)
=\gamma^{(S)}_{1\,N-1,n-1}(t-1)\quad (t>1)\,.
\label{dvNn2t}
\ee
Note that, since $V_{N,n}(v,1)=\delta_{v,1}$, the variance vanishes and the skewness remains undefined when $t=1$.

\subsection{Moments of $T_{N,n}(v,t)$}
The first derivative at $z=1$ of the generating function in~\eref{CTz} gives the mean value of the 
record times (covering times) as a function of the record value $v$:
\bea
\overline{t_{N,n}(v)}&=\sum_{t=1}^\infty t T_{N,n}(v,t)=\left.\frac{\partial\CT_{N,n}}{\partial z}\right|_{z=1}
=\left.\frac{\CT_{N,n}}{z}\frac{\partial\ln\CT_{N,n}}{\partial\ln z}\right|_{z=1}\nonumber\\
&=\left.\frac{\CT_{N,n}}{z}\sum_{k=0}^{v-1}\frac{N}{N-z(N-n+k)}\right|_{z=1}
=N(H_n-H_{n-v})\,.
\label{tNnv}
\eea
Here $H_n=\sum_{k=1}^n1/k$ is a harmonic number with $H_0=0$~\cite{graham94,flajolet09}.
The mean record time is the sum $\sum_{s=0}^{v-1}\overline{l_{N,n}(s)}$ of the mean lifetimes in~\eref{lNns}.
The second moment is given by
\bea
\fl\overline{t_{N,n}^2(v)}&=\sum_{t=1}^\infty t^2 T_{N,n}(v,t)=\left.\frac{\partial}{\partial z}
\left(z\,\frac{\partial\CT_{N,n}}{\partial z}\right)\right|_{z=1}
=\left.\frac{\partial}{\partial z}\left(\CT_{N,n}\sum_{k=0}^{v-1}\frac{N}{N-z(N-n+k)}\right)\right|_{z=1}\nonumber\\
\fl &=\left.\frac{\CT_{N,n}}{z}\left(\sum_{k=0}^{v-1}\frac{N}{N-z(N-n+k)}\right)^2\right|_{z=1}\!\!\!\!
+\left.\CT_{N,n}\sum_{k=0}^{v-1}\frac{N(N-n+k)}{[N-z(N-n+k)]^2}\right|_{z=1}\nonumber\\
\fl &=N^2\left[(H_n-H_{n-v})^2+H_{n,2}-H_{n-v,2}\right]-N(H_n-H_{n-v})\,,
\label{tNn2v}
\eea
where $H_{n,m}=\sum_{k=1}^n1/k^m$ is a generalized harmonic number with $H_{0,m}=0$~\cite{graham94,flajolet09}. 
Thus the variance takes the following form:
\be
\overline{\Delta t_{N,n}^2(v)}=N^2(H_{n,2}-H_{n-v,2})
-N(H_n-H_{n-v}).
\label{dtNn2v}
\ee
This is the sum $\sum_{s=0}^{v-1}\overline{\Delta l_{N,n}^2(v)}$ of the variances of the lifetimes in~\eref{dlNn2s}, 
as expected for a sum of independent random variables.
For the third moment we have:
\be
\overline{t_{N,n}^3(v)}=\sum_{t=1}^\infty t^3 T_{N,n}(v,t)=\left.\frac{\partial}{\partial z}\left[z\frac{\partial}{\partial z}
\left(z\,\frac{\partial\CT_{N,n}}{\partial z}\right)\right]\right|_{z=1}\,.
\label{tNn3v-1}
\ee

\begin{figure}[!th]
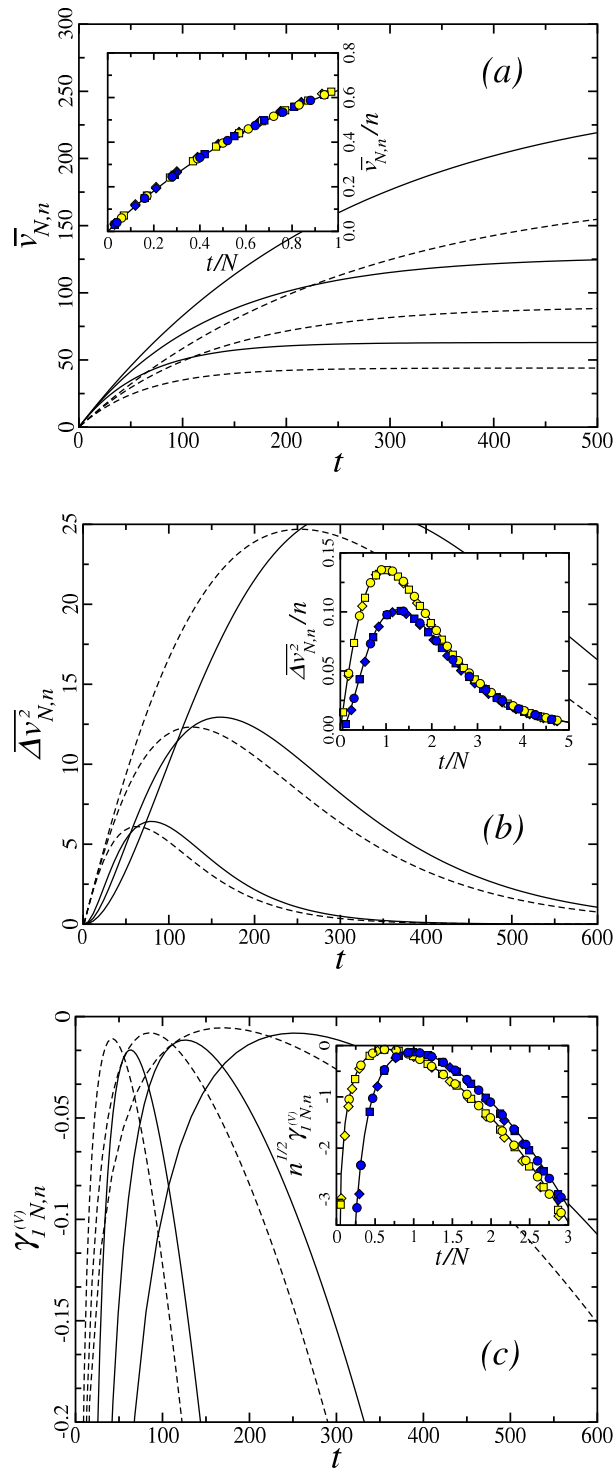

\begin{center}
\includegraphics[width=8cm,angle=0]{Fig-3a.eps}
\vglue 5mm
\includegraphics[width=8cm,angle=0]{Fig-3b.eps}
\vglue 5mm
\includegraphics[width=8cm,angle=0]{Fig-3c.eps}
\end{center}
\vglue -.5cm
\caption{The main panels give the time evolution for different lattice sizes of (a) the mean value 
$\overline{v_{N,n}(t)}$, (b) the variance $\overline{\Delta v^2_{N,n}(t)}$ and (c) the skewness 
$\gamma^{(V)}_{1\,N,n}(t)$ of the record value $v$ for the number of distinct sites visited by 
the random walk on the $n$-site sublattice. Full lines correspond to $f=n/N=1$ and dashed lines to $f=1/\sqrt{2}$.
Sizes are increasing from bottom to top or left to right in (c). The insets show the data collapse 
obtained with scaled variables (symbols) as well as the behaviour in the scaling limit (full line) 
as explained in section~4. Blue symbols correspond to $f=1$ and yellow symbols to $f=1/\sqrt{2}$. 
The sizes are the same as on the main panels with $N-n$ given by $64-64$ and $64-45$ (diamond), 
$128-128$ and $128-91$ (square), $256-256$ and $256-181$ (circle).
\label{fig-3}
}
\end{figure}
\clearpage

\noindent A lengthy but straightforward calculation leads to:
\bea
\fl\overline{t_{N,n}^3(v)}&=N^3\left[(H_n-H_{n-v})^3+3(H_n-H_{n-v})(H_{n,2}-H_{n-v,2})+2(H_{n,3}-H_{n-v,3})\right]\nonumber\\
\fl&\ \ \ \ \ \ -3N^2\left[(H_n-H_{n-v})^2+(H_{n,2}-H_{n-v,2})\right]+N(H_n-H_{n-v})\,.
\label{tNn3v-2}
\eea
Inserting these results in the expression of the skewness~\eref{gamX}, one obtains:
\be
\fl\gamma^{(T)}_{1\,N,n}(v)=\frac{1}{N^{1/2}}\,\frac{2N^2(H_{n,3}-H_{n-v,3})-3N(H_{n,2}-H_{n-v,2})+H_n-H_{n-v}}
{\left[N(H_{n,2}-H_{n-v,2})-H_n+H_{n-v}\right]^{3/2}}
\label{gamT}
\ee

\section{Moments and probability densities in the scaling limit}
In the scaling limit $N\to\infty$, $n\to\infty$, $t\to\infty$ or $v,s\to\infty$, the appropriate 
combinations of variables are~\footnote[2]{See I for a discussion of the finite-size scaling 
behaviour on the fully-connected lattice}
\be
\frac{n}{N}\scal f\,,\qquad \frac{t}{N}\scal w\,,\qquad \frac{v}{n}\scal x\,,\qquad \frac{s}{n}\scal x'\,.
\label{scal}
\ee
In this section we look for the behaviour of the probability distributions in this limit.

\subsection{Records lifetime}
Using scaled variables, the mean value of the lifetime in~\eref{lNns} and its variance in~\eref{dlNn2s} take the following form:
\be
\overline{l_{N,n}(s)}\scal\frac{1}{f(1-x')}\,,\qquad\overline{\Delta l^2_{N,n}(s)}
=\scal\left[\frac{1}{f(1-x')}\right]^2-\frac{1}{f(1-x')}\,.
\label{scall}
\ee
Note that they are diverging as total covering is approached, i.e., when $x'\to 1$.
The probability distribution of the lifetimes in~\eref{Jsl}, which is exponential in $l$, takes the following form 
\be
J_{N,n}(s,l)\simscal f(1-x')\,\e^{-f(1-x')\,l}\,,
\label{scalJsl}
\ee
close to total covering.

\subsection{Number of records and records values}
According to~\eref{Vvt-2} the probability distributions $S_{N,n}(s,t)$ and $V_{N,n}(v,t)$ have 
the same scaling limit. There first moment
follows from~\eref{sNnt} and~\eref{vNnt} and scales as $n$ so that:
\be
\frac{\overline{s_{N,n}(t)}}{n}\scal\frac{\overline{v_{N,n}(t)}}{n}\scal 1-\e^{-w}\,,\qquad w=t/N\,.
\label{scalsv}
\ee
The mean value grows initially as $ft$ (infinite system behaviour). The approach to the saturation 
value, $n$, is exponential, with a relaxation time equal to $N$ (see figure~\ref{fig-3}(a)). 
The inset shows a good data collapse on the scaling function~\eref{scalsv}.

The variance given, by~\eref{dsNn2t} and~\eref{dvNn2t}, scales as $n$ too and behaves as:
\be
\frac{\overline{\Delta s^2_{N,n}(t)}}{n}\scal\frac{\overline{\Delta v^2_{N,n}(t)}}{n}\scal\phi(f,w)=\e^{-w}-(1+fw)\e^{-2w}\,.
\label{scalds2dv2}
\ee
Although the variance is extensive in the scaling limit, we show in appendix B  that its value per site
remains modified by correlations. The initial growth is linear when $f<1$, quadratic when $f=1$ and the long-time decay is exponential 
(see figure~\ref{fig-3}(b)). The fluctuations are at their maximum for~$t$ close to~$N$ 
($w_{\mathrm {max}}=1.256431\ldots$ for $f=1$ and $w_{\mathrm {max}}=0.991353\ldots$ 
for $f=1/\sqrt{2}$). The inset shows the data collapse on two $f$-dependent scaling functions given by~\eref{scalds2dv2}.

For the skewness in~\eref{gamS} and~\eref{dvNn2t}, vanishing as $n^{-1/2}$, one obtains:
\be
\fl n^{1/2}\gamma^{(S,V)}_{1\,N,n}(t)\scal
\frac{-\e^{-w}+\!3(1+fw)\e^{-2w}-(2+\!6fw-2f^2w+3f^2w^2)\e^{-3w}}{\left[\e^{-w}-(1+fw)\e^{-2w}\right]^{3/2}}.
\label{scalgsv}
\ee
When $w\ll1$ the leading contribution to~\eref{scalgsv} is such that
\be
n^{1/2}\gamma^{(S,V)}_{1\,N,n}(t)\scal-\frac{2f-1}{[w(1-f)]^{1/2}}\qquad (f<1)\,,
\label{scalgsvl1}
\ee
and
\be
n^{1/2}\gamma^{(S,V)}_{1\,N,N}(t)\scal-\frac{\sqrt{2}}{w}\qquad (f=1)\,.
\label{scalgsvl2}
\ee
When $w\gg1$ an expansion in powers of $\e^{-w}$ gives:
\be
n^{1/2}\gamma^{(S,V)}_{1\,N,n}(t)\scal-\e^{w/2}\left[1-\frac{3}{2}(1+fw)\e^{-w}+{\mathrm O}\!\left(\e^{-2w}\right)\right]\,.
\label{scalgsvg}
\ee
The skewness in figure~\ref{fig-3}(c) goes through a negative maximum value for $t$ close to $n$. It is 
diverging in the limits $w\to0$ and $w\to\infty$ but this does not mean an increase of the asymmetry. 
On the contrary a closer look at~\eref{scalgsv} shows that the numerator, given by the third centered 
moment and measuring the asymmetry of the distribution, is vanishing as $w$ when $f<1$ and $w^2$ 
when $f=1$ for $w\ll1$ and as $\e^{-w}$ for $w\gg1$. Thus the divergences are due to the cube of the 
standard deviation in the denominator, which vanishes more rapidly in these limits. The inset shows 
the data collapse on the two scaling functions following from~\eref{scalgsv}. 

Let us now determine the common scaling limit of the probability distributions $S_{N,n}(s,t)$ and $V_{N,n}(v,t)$.
We shall do it starting from the master equation~\eref{masteq-1}. The scaling behaviour of the variance in~\eref{scalds2dv2}
suggests the introduction of the dimensionless variable:
\be
\frac{s-\overline{s_{N,n}(t)}}{n^{1/2}}\scal\sigma=\frac{s}{n^{1/2}}-n^{1/2}(1-\e^{-w})\,.
\label{scals}
\ee
A properly normalized probability density $\fS_(\sigma,f,w)$ is given by $n^{1/2}S_{N,n}(s,t)$ in the scaling limit.
Then as shown in appendix C, to leading order in an expansion in powers of $N^{-1/2}$, the master equation~\eref{masteq-1} 
leads to the following partial differential equation for the probability density:
\be
\frac{\partial\fS}{\partial w}=\frac{\e^{-w}}{2} (1-f\e^{-w})\frac{\partial^2\fS}{\partial \sigma^2}+\sigma\frac{\partial\fS}{\partial \sigma}+\fS\,.
\label{pdfS-1}
\ee
Rewriting the probability density as
\be
\fS(\sigma,f,w)=\fP[\sigma,\phi(f,w)]\,,
\label{SP-1}
\ee
where $\phi(f,w)$ is the scaling function defined in~\eref{scalds2dv2} one has
\be\fl
\e^{-w}(1-f\e^{-w})=2\phi+\frac{\partial\phi}{\partial w}\,,\qquad
\frac{\partial\fS}{\partial w}\!=\!\frac{\partial\fP}{\partial\phi}\frac{\partial\phi}{\partial w}\,,\qquad
\frac{\partial\fS}{\partial \sigma}\!=\!\frac{\partial\fP}{\partial \sigma}\,,\qquad
\frac{\partial^2\fS}{\partial \sigma^2}\!=\!\frac{\partial^2\fP}{\partial \sigma^2}\,,
\label{SP-2}
\ee
and the partial differential equation~\eref{pdfS-1} can be rewritten as:
\be
\frac{\partial\phi}{\partial w}\left(\frac{\partial\fP}{\partial\phi}-\frac{1}{2}\frac{\partial^2\fP}{\partial \sigma^2}\right)=
\phi\frac{\partial^2\fP}{\partial \sigma^2}+\sigma\frac{\partial\fP}{\partial \sigma}+\fP\,.
\label{pdfP}
\ee
It is easy to verify that with the Gaussian density 
\be
\fP(\sigma,\phi)=\frac{\e^{-\sigma^2/(2\phi)}}{\sqrt{2\pi\phi}}\,,\qquad\phi(f,w)=\e^{-w}-(1+fw)\e^{-2w}\,,
\label{fP}
\ee
which is a solution of the diffusion equation on the left, the right-hand side vanishes too.

Starting from the master equation
\be\fl
V_{N,n}(v,t)=\frac{N-n+v-1}{N-1}V_{N,n}(v,t-1)+\frac{n-v+1}{N-1}V_{N,n}(v-1,t-1)\,, \quad t>1\,,
\label{masteq-2}
\ee
which follows from~\eref{masteq-1} and~\eref{Vvt-2},
the same Gaussian solution~\eref{fP} is obtained for the probability density of the records values 
with $s$ replaced by $v$ in the expression~\eref{scals} of the scaling variable $\sigma$. 

The scaling behaviour is shown in figure~\ref{fig-4}(a) for $f=1$ and $f=1/\sqrt{2}$ at three values of $w$.
A mean-field calculation is given in appendix B.

\subsection{Records times at partial covering}
Let us first consider the case of records times corresponding to a partial covering of the sublattice, $x<1$ ($n-v={\mathrm O}(n)$).
For large values of $k$ the harmonic numbers can be written as
\be
H_k=\ln k+\gamma+{\mathrm O}(1/k)\,,
\label{Hn}
\ee
where $\gamma$ is the Euler--Mascheroni constant. 
Thus in the scaling limit we have
\be
H_n-H_{n-v}\scal\ln n-\ln(n-v)=-\ln\left(1-\frac{v}{n}\right)=-\ln(1-x)\,,
\label{Hnv}
\ee
and the behaviour of the partial covering time follows from~\eref{tNnv} with:
\be
\frac{\overline{t_{N,n}(v)}}{N}\scal-\ln(1-x)\qquad (x<1)\,.
\label{scalt-1}
\ee
Note that this expression leads to
\be
\frac{v}{n}\scal 1-\exp\left(-\frac{\overline{t_{N,n}(v)}}{N}\right)_{\mathrm s.l.}\,,
\label{invscalt-1}
\ee
to be compared to~\eref{scalsv}. The mean value $\overline{t_{N,n}(v)}$, shown in~figure~\ref{fig-5}(a), 
initially grows as $v/f$. The inset illustrates the collapse of the finite-size data on a single scaling function.

\begin{figure}[!t]
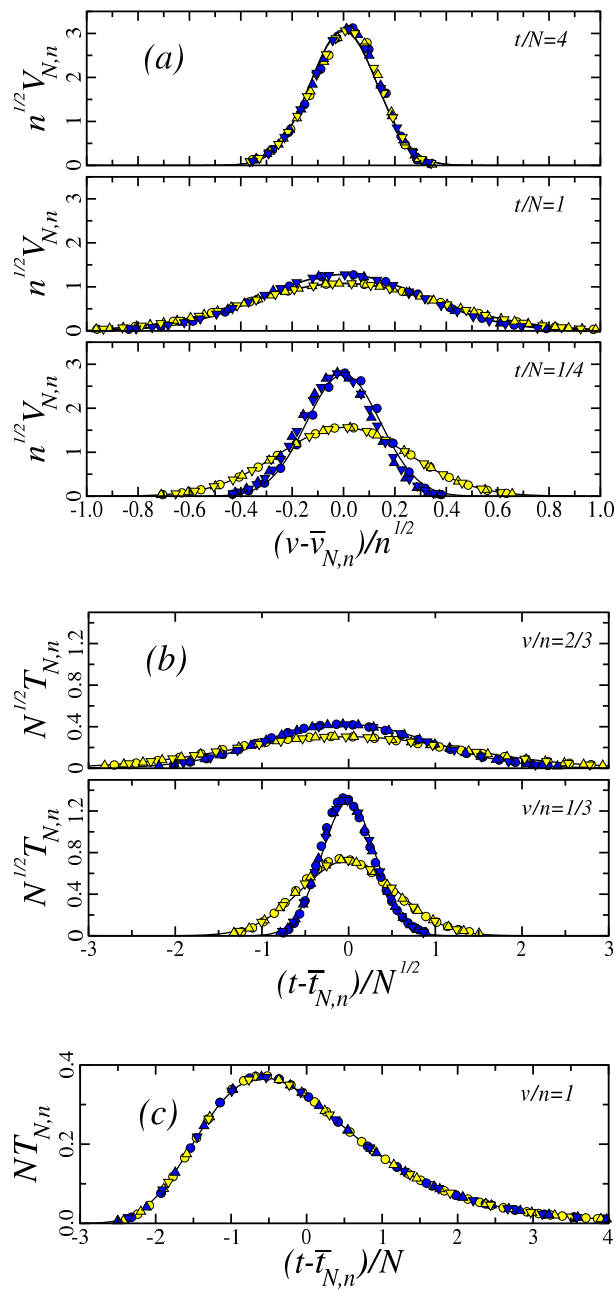

\begin{center}
\includegraphics[width=8cm,angle=0]{Fig-4a.eps}
\vglue 5mm
\includegraphics[width=8cm,angle=0]{Fig-4b.eps}
\vglue 5mm
\includegraphics[width=8cm,angle=0]{Fig-4c.eps}
\end{center}
\vglue -.5cm
\caption{
\label{fig-4} Data collapse obtained for (a) the scaled probability distribution $n^{1/2}V_{N,n}(v,t)$ as a function of
 $(v-\overline{v_{N,n}(t)})/n^{1/2}\scal\sigma$ for different values of $t/N\scal w$, (b) the scaled probability distribution 
 $N^{1/2}T_{N,n}(v,t)$ as a function of $(t-\overline{t_{N,n}(v)})/N^{1/2}\scal\tau$ for different values of $v/n\scal x$ 
 and (c) the scaled probability distribution $NT_{N,n}(v,t)$ as a function of $(t-\overline{t_{N,n}(v)})/N\scal\tau'$. 
 Blue symbols correspond to $f=1$ and yellow symbols to $f=1/\sqrt{2}$. The sizes $N-n$ are given by $256-256$ 
 and $256-181$ (circle), $512-512$ and $512-362$ (triangle up), $1024-1024$ and $1024-724$ (triangle down). 
 The full lines give the probability densities obtained in the scaling limit: (a) the Gaussian density $\fP[\sigma,\phi(f,w)]$ 
 given by~\protect\eref{fP}, (b) the Gaussian density $\fQ[\tau,\chi(f,x)]$ given by~\protect\eref{fQ} 
 and (c) the type-I Gumbel distribution $\fT'_0(\tau')$ given by~\protect\eref{gumbel0}. 

}
\end{figure}

\begin{figure}[!th]
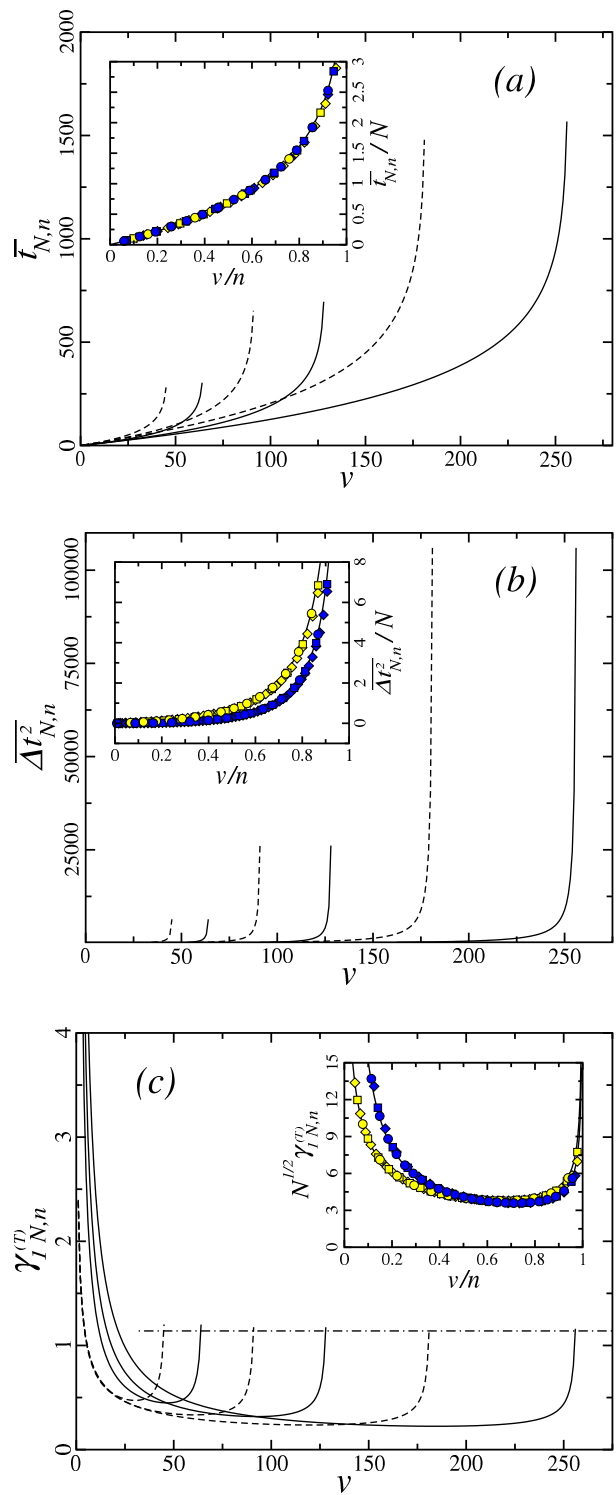

\begin{center}
\includegraphics[width=8cm,angle=0]{Fig-5a.eps}
\vglue 5mm
\includegraphics[width=8cm,angle=0]{Fig-5b.eps}
\vglue 5mm
\includegraphics[width=8cm,angle=0]{Fig-5c.eps}
\end{center}
\vglue -.5cm
\caption{For different lattice sizes, the main panels give the evolution with $v$ of (a) the mean value 
$\overline{t_{N,n}(v)}$, (b) the variance $\overline{\Delta t^2_{N,n}(v)}$ and (c) the skewness 
$\gamma^{(T)}_{1\,N,n}(v)$ of the record time $t$ needed by the random walk to visit $v$ distinct 
sites on the $n$-site sublattice. Full lines correspond to $f=n/N=1$ and dashed lines to $f=1/\sqrt{2}$.
Sizes are increasing from left to right. The dashed-dotted line indicates the value of the skewness for $v=n$ 
in the scaling limit (see equation~\protect\eref{scalgt-2} with u=0). The insets show the data collapse obtained with scaled 
variables (symbols) as well as the behaviour in the scaling limit (full line) as explained in section~4. 
Blue symbols correspond to $f=1$ and yellow symbols to $f=1/\sqrt{2}$. The sizes are the same as 
on the main panels with $N-n$ given by $64-64$ and $64-45$ (diamond), $128-128$ and $128-91$ (square), 
$256-256$ and $256-181$ (circle).
\label{fig-5}
}
\end{figure}
\clearpage

Differences involving generalized harmonic numbers in~\eref{dtNn2v} and~\eref{gamT} 
can be evaluated in the scaling limit as follows:
\be\fl
H_{n,m}-H_{n-v,m}=\sum_{k=n-v+1}^n\frac{1}{k^m}\scal\int_{n-v}^n\frac{du}{u^m}
=\frac{1}{m-1}\left[\frac{1}{(n-v)^{m-1}}-\frac{1}{n^{m-1}}\right]\,.
\label{Hnvm}
\ee
Using~\eref{Hnv} and~\eref{Hnvm} in the expression of the variance~\eref{dtNn2v}, one obtains:
\be
\frac{\overline{\Delta t_{N,n}^2(v)}}{N}\scal\chi(f,x)=\frac{x}{f(1-x)}+\ln(1-x)\qquad (x<1)\,.
\label{scaldt2-1}
\ee
The initial growth of the variance, shown in~figure~\ref{fig-5}(b), is linear when $f<1$ and quadratic 
when $f=1$. The finite-size data collapse on the scaling functions is illustrated in the inset.

When $x<1$ the skewness in~\eref{gamT} behaves as:
\be
\fl N^{1/2}\gamma^{(T)}_{1\,N,n}(v)\!\scal\!\left[\frac{x(2\!-\!x)}{f^2(1\!-\!x)^2}
-\frac{3x}{f(1\!-\!x)}-\ln(1\!-\!x)\right]\left/\left[\frac{x}{f(1\!-\!x)}+\ln(1\!-\!x)\right]^{3/2}\right.\!\!\!\!\!.
\label{scalgt-1}
\ee
When $x\to0$, to leading order, the last equation leads to
\be
N^{1/2}\gamma^{(T)}_{1\,N,n}(v)\scal\frac{2xf^{-2}
-3xf^{-1}+x}{\left(xf^{-1}-x\right)^{3/2}}\scal\frac{2-f}{[xf(1-f)]^{1/2}}\qquad (f<1)\,,
\label{scalgtx-1}
\ee
whereas
\be
N^{1/2}\gamma^{(T)}_{1\,N,N}(v)\scal\frac{\sqrt{2}}{x}\qquad (f=1)\,.
\label{scalgtx-2}
\ee
When $x\to1$, to leading order, one obtains:
\be
N^{1/2}\gamma^{(T)}_{1\,N,n}(v)\scal\frac{1}{[f(1-x)]^{1/2}}\qquad (x<1)\,.
\label{scalgtx-3}
\ee
The skewness is shown as a function of $v$ in figure~\ref{fig-5}(c). The inset illustrates the data collapse 
on the  two scaling functions following from~\eref{scalgt-1}.
The scaled skewness is diverging in both limits but for different reasons. 
The numerator and the denominator of~\eref{scalgt-1} vanish when $x\to0$ and diverge when $x\to1$. 
Thus the divergence is governed by the vanishing variance when $x\to0$ and by
the third centered moment (a measure of the asymmetry) when $x\to1$. 

Let us now consider the behaviour of the probability distribution $T_{N,n}(v,t)$ in the scaling limit. 
Using~\eref{Tvt-2} in~\eref{masteq-1} one obtains the following master equation:
\be\fl
T_{N,n}(v,t)=\frac{N-n+v-1}{N}T_{N,n}(v,t-1)+\frac{n-v+1}{N}T_{N,n}(v-1,t-1)\,, \quad t>1\,,
\label{masteq-3}
\ee
Taking into account the scaling behaviour of the variance in~\eref{scaldt2-1} the following dimensionless variable is appropriate:
\be
\frac{t-\overline{t_{N,n}(v)}}{N^{1/2}}\scal\tau=\frac{t}{N^{1/2}}+N^{1/2}\ln(1-x)\,.
\label{scalt}
\ee
Thus the normalized probability density $\fT(\tau,f,x)$ is obtained as the continuum limit 
of $N^{1/2}T_{N,n}(v,t)$. As shown in appendix D, to leading order in an expansion in powers 
of $N^{-1/2}$, it satisfies the relatively simple partial differential equation:
\be
\frac{\partial\fT}{\partial x}=\frac{1-f(1-x)}{2f(1-x)^2}\frac{\partial^2\fT}{\partial\tau^2}\,.
\label{pdfT-1}
\ee
Noticing that $\chi(f,x)$ in~\eref{scaldt2-1} is such that
\be
\frac{\partial\chi}{\partial x}=\frac{1-f(1-x)}{f(1-x)^2}\,,
\label{chix}
\ee
one can use the change of variables $\fT(\tau,f,x)=\fQ[\tau,\chi(f,x)]$
to further simplify~\eref{pdfT-1}. Doing so, the diffusion equation is finally obtained:
\be
\frac{\partial\fQ}{\partial\chi}=\frac{1}{2}\frac{\partial^2\fQ}{\partial \tau^2}\,.
\label{pdfQ}
\ee
Thus the probability density is Gaussian in the scaling limit and given by:
\be
\fQ(\tau,\chi)=\frac{\e^{-\tau^2/(2\chi)}}{\sqrt{2\pi\chi}}\,,\qquad\chi(f,x)=\frac{x}{f(1-x)}+\ln(1-x)\,.
\label{fQ}
\ee
Note that $\chi(f,x)$, the scaled variance $\overline{\Delta t_{N,n}^2(v)}/N$ in the scaling limit, 
is diverging at $x=1$. This means that a new scaling variable is needed at almost total and total covering.
The finite-size data collapse on the Gaussian densities is shown in figure~\ref{fig-4}(b).

\subsection{Records times at almost total and total covering}
Let us now study the vicinity of total covering, i.e., the records times at $v$ such that the number of 
unvisited sites $u=n-v={\mathrm O}(1)$. 
Then $u/n\scal1-x\scal0$ in the scaling limit, hence the scaling functions in~\eref{scalt-1},~\eref{scaldt2-1} and~\eref{scalgt-1} diverge: 
the scaling behaviour is anomalous. When $n$ is large, inserting~\eref{Hn} into~\eref{tNnv} gives the mean value: 
\be
\overline{t_{N,n}(v)}\simeq N(\ln n+\gamma-H_u)\,,\qquad u=n-v\,.
\label{scalt-2}
\ee
One may write
\be
H_{n,m}=\!\sum_{k=1,n}\frac{1}{k^m}\simeq\zeta(m)-\int_n^\infty\!\frac{du}{u^m}
\simeq\zeta(m)-\frac{n^{1-m}}{m\!-\!1}\qquad(m>1)\,,
\label{Hnm}
\ee
where $\zeta(m)=\sum_{k=1}^\infty 1/k^m$ is the Riemann zeta function. Making use of~\eref{Hn} and~\eref{Hnm} 
with $m=2$ in~\eref{dtNn2v}, one obtains the variance:
\be
\frac{\overline{\Delta t_{N,n}^2(v)}}{N^2}\scal\zeta(2)-H_{u,2}\,.
\label{scaldt2-2}
\ee
Thus, when $u=n-v={\mathrm O}(1)$ the variance is independent of $f$ in the scaling limit and scales as $N^2$ instead of $N$ for $x<1$.
The behaviour of the skewness in~\eref{gamT} is obtained using~\eref{Hn} and~\eref{Hnm} and reads:
\be
\gamma^{(T)}_{1\,N,n}(v)\scal\frac{2[\zeta(3)-H_{u,3}]}{[\zeta(2)-H_{u,2}]^{3/2}}\,.
\label{scalgt-2}
\ee

\begin{figure}[!t]
\begin{center}
\includegraphics[width=8cm,angle=0]{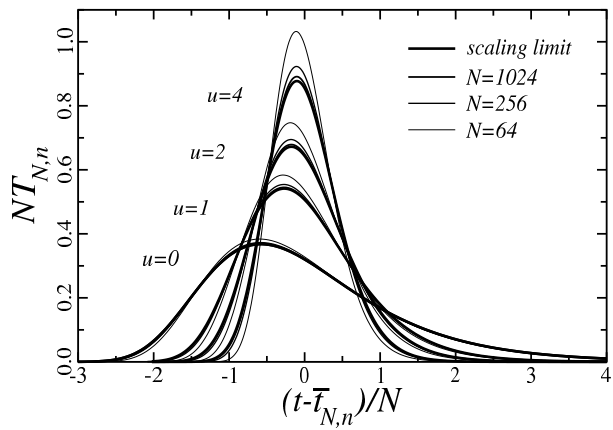}
\end{center}
\vglue -.5cm
\caption{Evolution of the covering time probability distribution in the immediate vicinity of total covering. In the scaling limit (thick lines) the standard type-I Gumbel distribution ($u=n-v=0$) is approached 
via a sequence of generalized Gumbel distributions, indexed by the deviation from total covering $u=4,2,1$. 
Thinner lines show finite-size effects for $N=64$, $256$ and $1024$ from top to bottom.
\label{fig-6} 

}
\end{figure}

According to~\eref{scaldt2-2}, the appropriate scaling variable for the probability density is now
\be
\frac{t-\overline{t_{N,n}(v)}}{N}\simeq\tau'=\frac{t}{N}-\ln n-\gamma+H_u\,,\qquad u=n-v\,.
\label{scalt'}
\ee
As a consequence, the probability density $\fT'_u(\tau')$ is given by the scaling limit of~$NT_{N,n}(n-u,t)$ which, 
as shown in appendix E using the properties of Stirling numbers, can be written as~\footnote[3]{Note that 
when $n=N$ equation~\eref{NTNn-2} with $u=0$ is in agreement with the probability distribution obtained in~\cite{zlatanov09} 
for the complete graph up to the substitutions $N\to N-1$ and $t-1\to t$ needed due to different step rules 
and time origins. It also agrees with the conjecture formulated earlier in~\cite{nemirovsky91}.}:
\be\fl
NT_{N,n}(n-u,t)=\frac{n!}{u!\,(n-u-1)!}\sum_{l=0}^{n-u-1}(-1)^{l}{n-u-1\choose l}\left(1-\frac{l+u+1}{N}\right)^{t-1}\!\!\!.
\label{NTNn-2}
\ee
In the scaling limit one may write
\be
\left(1-\frac{l+u+1}{N}\right)^{t-1}\simeq \e^{-(l+u+1)t/N}
\simeq\frac{\e^{-(l+u+1)(\tau'+\gamma-H_u)}}{n^{l+u+1}}\,,
\label{exp}
\ee
where the last equality follows from~\eref{scalt'}. To leading order in $n$, one has
\be
\frac{n!}{(n-u-1)!}{n-u-1\choose l}=\frac{n!}{(n-l-u-1)!\,l!}\simeq\frac{n^{l+u+1}}{l!}\,.
\label{choose}
\ee
Thus~\eref{NTNn-2} leads to
\bea
\fT'_u(\tau')&=\frac{\e^{-(u+1)(\tau'+\gamma-H_u)}}{u!}\sum_{l=0}^\infty(-1)^l\,\frac{\e^{-l(\tau'\!+\gamma-H_u)}}{l!}\nonumber\\
&=\frac{1}{u!}\exp\left[\!-(u+1)(\tau'\!+\gamma\!-H_u)-\e^{-(\tau'\!+\gamma-H_u)}\right]\,,
\label{gumbel}
\eea
in the scaling limit. $\fT'_u(\tau')$ is a generalized Gumbel distribution~\cite{ojo01,pinheiro15} which 
gives the standard type-I Gumbel distribution~\cite{gumbel35,gumbel04} 
\be
\fT'_0(\tau')=\exp\left[-(\tau'\!+\gamma)-\e^{-(\tau'+\gamma)}\right]\,,
\label{gumbel0}
\ee
when $u=0$, i.e., at total covering (see figure~\ref{fig-4}(c)). The discrete evolution of the probability distribution as total covering is approached ($u=n-v={\mathrm O}(1)$) is shown in figure~\ref{fig-6}.

\subsection{Crossover from Gumbel to Gauss}
In this section we study the crossover of the probability distribution~\eref{gumbel} to Gaussian behaviour, when the deviation from total covering becomes larger, with $n\gg u\gg1$.
The generalized Gumbel distribution can be written as:
\be
\fT'_u(\tau')=\exp\left[f_{u+1}(\tau'+\gamma-H_u)\right]\,,\quad f_a(b)=-ab-\e^{-b}-\ln\Gamma(a)\,.
\label{fab}
\ee
Let us expand $f_a(b)$ to second order in the vicinity of its maximum at $b_0$:
\be
f_a(b)\simeq f_a(b_0)+\frac{1}{2}f_a''(b_0)(b-b_0)^2
\label{fab1}
\ee
One has:
\be
f_a'(b)=-a+\e^{-b}\,,\quad b_0=-\ln a\,,\quad f_a''(b)=-\e^{-b}\,,\quad f_a''(b_0)=-a\,.
\label{fab3}
\ee
When $a\gg1$ the logarithm of the Gamma function has the following expansion~\cite{abramowitz72}:
\be
\ln\Gamma(a)\simeq a \ln a-a +\frac{1}{2}\ln\left(\frac{2\pi}{a}\right)\,.
\label{lnG}
\ee
Finally:
\be
f_a(b)\simeq -\frac{1}{2}\ln\left(\frac{2\pi}{a}\right)-\frac{a}{2}(b+\ln a)^2\,.
\label{fab4}
\ee
With $b=\tau'+\gamma-H_u$, $a=u+1\simeq u$ in~\eref{fab}, using~\eref{Hn} one obtains:
\be
b+\ln a\simeq \tau'+\gamma-H_u+\ln u\simeq \tau'\,.
\label{blna}
\ee
Thus, according to~\eref{fab4}, the generalized Gumbel distribution in~\eref{fab} can be approximated by
\be
\fT'_u(\tau')\simeq \frac{\e^{-u{\tau'}^2/2}}{\sqrt{2\pi/u}}\,,
\label{gumgau1}
\ee
when $n\gg u\gg1$ in the vicinity of $\tau'=0$.
In order to compare to the Gaussian distribution $\fT(\tau,f,x)=\fQ(\tau,\chi)$ in~\eref{fQ},
obtained for $x<1$ in the scaling limit, one has to change the scaling variable 
from~$\tau'$ in~\eref{scalt'} to~$\tau$~in\eref{scalt} so that:
\be
\frac{t-\overline{t_{N,n}(v)}}{N}\scal\tau'=\frac{\tau}{\sqrt{N}}\,.
\label{tau'tau}
\ee
When $N$ is large, the behaviour of the probability distribution as a function of $\tau$ is 
governed by the immediate vicinity of $\tau'=0$. This justifies {\it a posteriori} the second order expansion 
used above. The change of variable leads to the following Gaussian distribution:
\be
\fT_u(\tau)\simeq\frac{\fT'_u(\tau')}{\sqrt{N}}=\sqrt{\frac{u}{2\pi N}}\exp\left(-\frac{u\tau^2}{2N}\right)\,.
\label{gumgau2}
\ee
This expression has to be compared to~\eref{fQ} when $x$ is close to $1$. In this limit, 
the logarithmic contribution to $\chi(f,x)$ can be neglected so that
\be
\chi(f,x)\simeq\frac{1}{f(1-x)}=\frac{N}{n-v}=\frac{N}{u}\,,
\label{chi1}
\ee
and a perfect agreement between the two distributions is obtained.

\section{Discussion and outlook}
As mentioned in I, finite-size scaling results obtained for the fully-connected lattice ($d=\infty$) 
are expected to be representative of the behaviour on periodic lattices above the critical 
dimension $d_{\mathrm c}=2$.

The effect of the restriction to a sublattice with $n$ sites on the mean number of distinct sites visited 
$\overline{s_{N,n}(t)}$ given in~\eref{sNnt} is quite simple. When normalized by the mean value of 
the total number of visits of the sublattice, $tn/N$, the ratio is 
independent of $n$ and thus the same as for the full lattice. This property remains valid for periodic 
lattices in $d=1$ to 3 as shown in~\cite{weiss82}. 

A generalized covering time for random walks on graphs, the marking time, has been introduced in~\cite{banderier00}. 
The walker marks a visited site $i$ with probability $p_i$ and the marking time is the time needed to mark all the sites.
On the fully connected lattice, when the marking probability is uniform and equal to $p$, it is easy to verify 
that $N$ has simply to be changed into $N/p$ in the generating function~\eref{CTz}. Thus, according 
to~\eref{tNnv}, the total marking time is given by $NH_n/p$. 

The probability distribution for the number of distinct sites $s$ visited up to time $t$ in 
the subset (record number at $t$) and the probability distribution for the record value at a given 
record time lead to the same Gaussian density in the scaling limit. This is 
the probability density obtained in the mean-field approximation when correlations 
between the visits of different sites are neglected (see appendix B). But there is a remnant of the correlation
effects in the variance which differs from the mean-field result by a term of order $\e^{-2w}$.
We have also obtained a Gaussian density for the record times at partial covering in the scaling limit. 
Finite-size corrections to the partial differential equations leading to these results
are of relative order $N^{-1/2}$. Consistently, the asymmetry of the corresponding discrete distributions, 
as measured by the skewness, also vanishes as $N^{-1/2}$. 

The record times are given by the sum of $v$ independent, differently distributed random variables, 
which are the lifetimes of previous records. Their variance scales as $N$ at partial covering and as 
$N^2$ at almost total and total covering. Hence in the scaling variables, $\tau$ for $x<1$ and $\tau'$ 
for $x=1$, the time has to be divided  
by $N^{1/2}$ and $N$, respectively. The probability distribution of the record times is Gaussian at 
partial covering, as expected since the central limit theorem applies for a sum of independent 
random variables  with finite variances. 
The generalized Gumbel distribution at almost total covering  and the type-I Gumbel distribution 
at total covering are linked to the divergence of the variance of the 
lifetimes as $s/n\scal x'\to 1$ (see \eref{scall}). The main contribution to the record 
time then comes from lifetimes of records with number $s$ close to $n$, with an exponential distribution given in~\eref{scalJsl}. Note that this is not a standard case for Gumbel statistics which is usually associated with the distribution of extremes in a collection of random variables.

The generalized Gumbel distribution, indexed by the deviation $u=n-v$ from total covering, is crossing over 
to the Gaussian distribution obtained at partial covering when $u$ increases. At finite size, the crossover  between 
the two regimes can be observed on the mean value, the variance and the skewness in figure~\ref{fig-5}. 
The skewness in figure~\ref{fig-5}(c) increases when $v$ goes to $n$, reaching a value which converges 
from above to that obtained at $v=n$ in the scaling limit.

The time evolution of $s$, the number of distinct sites visited by the walker on the sublattice with $n$ 
sites, may be reinterpreted in different ways. Let us mention the example of a directed random walk 
with waiting times in $1D$ where $s$ in figure~\ref{fig-2}(a) gives the position of the 
walker on the segment $[0,n]$ as a function of time. The walker at $s$ either takes a step forward 
with probability $p(s)=(n-s)/N$ or waits with probability $1-p(s)$. The waiting times correspond 
to the lifetimes of the records. They are distributed according to $J_{N,n}(s,l)$ in~\eref{Jsl} 
and depend on the position of the walker. The mean velocity at $s$ is $p(s)$ so that the walker 
is slowing down as $s$ increases and stops at $s=n$. In this interpretation the partial 
covering time is the first-passage time at an intermediate point $s=v<n$ whereas the total covering 
time is the time of arrival at $s=n$.

The random walk problem on the fully-connected lattice has a Russian dolls generalization 
with $\nu$ walkers. The first walker performs a random walk on the fully connected lattice 
with $N$ sites. The second is only allowed to take a step on the sites previously visited by the first, 
and so on and so forth. We are currently studying the $1D$ transcription with a queue 
involving $\nu$ random walkers in interaction on the segment $[0,N]$. The walks are directed 
with waiting times depending of the distance between successive walkers. A walker is slowing down 
when approaching the preceding one in such a way that the initial order is always 
preserved~\footnote[3]{Note that in the recent calculation of the exact probability distribution 
for the number of distinct and common sites visited by $\nu$ walkers in $d=1$~\cite{kundu13,majumdar12b} 
the walks are random and independent whereas they are constrained in our case.}.

\ack
Helpful comments from Maur{\'\i}cio D. Coutinho-Filho, Jean-Yves Fortin and two anonymous referees are gratefully acknowledged.

\noindent {\sl Note added in proof.} It has recently been shown~\cite{chupeau15} that the generalized Gumbel distribution in \eref{gumbel} is universal for Markovian, non-compact random walks.

\appendix

\section{Bivariate generating function for $\bi{S_{N,n}(s,t)}$}
Let us define a bivariate generating function $\CS_{N,n}(y,z)$ which is ordinary in $y$ and exponential in $z$:
\be\fl
\CS_{N,n}(y,z)=\sum_{s=0}^\infty y^s\sum_{t=0}^\infty \frac{z^t}{t!}S_{N,n}(s,t)
=\sum_{s=0}^n n^{\underline{s}}\,y^s\sum_{t=0}^\infty \frac{(z/N)^t}{t!}{N-n+t\brace N-n+s}_{N-n}\,.
\label{CSyz-1}
\ee
A Stirling number of the second kind can be expressed as (see I, equation~(2.11))
\be
{k\brace s}=\frac{1}{s!}\left.\BD^s \eta^k\right|_{\eta=0}\,,
\label{Stir}
\ee
where $\BD$ is the forward-difference operator such that $\BD f(\eta)=f(\eta+1)-f(\eta)$. 
Making use of the definition~\eref{rStir-2} for a $r$-Stirling number of the second kind, one obtains
\be\fl
{r+t\brace r+s}_r=\sum_{k=0}^t{t\choose k}{k\brace s}\,r^{t-k}=\left.\frac{1}{s!}\BD^s
\sum_{k=0}^t{t\choose k}\,\eta^kr^{t-k}\right|_{\eta=0}=\left.\frac{1}{s!}\BD^s(\eta+r)^t\right|_{\eta=0}\,.
\label{rStir-4}
\ee
Thus the generating function may be rewritten as:
\bea
\CS_{N,n}(y,z)&=\sum_{s=0}^n \frac{n^{\underline{s}}\,y^s}{s!}
\left.\BD^s\sum_{t=0}^\infty \frac{(z/N)^t}{t!}(\eta+N-n)^t\right|_{\eta=0}\nonumber\\
&=\e^{z(1-n/N)}\sum_{s=0}^n{n\choose s}y^s\left.\BD^s\e^{z\eta/N}\right|_{\eta=0}\,,
\label{CSyz-2}
\eea
Since the result of the action of $\BD$ on $\e^{z\eta/N}$ is a multiplication by $(\e^{z/N}-1)$, one finally obtains:
\be\fl
\CS_{N,n}(y,z)=\e^{z(1-n/N)}\sum_{s=0}^n{n\choose s}\left[y(\e^{z/N}-1)\right]^s
=\e^{z(1-n/N)}\left[1+y(\e^{z/N}-1)\right]^n\,.
\label{CSyz-3}
\ee
When $n=N$ equation~(2.13) of I is recovered.

\section{Mean-field approximation for $S_{N,n}(s,t)$}
The probability that a given site has never been visited up to time $t$ is given by:
\be
q_N(t)=\left(\frac{N-1}{N}\right)^t\,.
\label{qNt}
\ee
Hence in the mean-field approximation, neglecting correlations between the visits of different sites,
the probability distribution for the number of sites visited on the subset 
up to times $t$ is the binomial distribution:
\be
S'_{N,n}(s,t)={n\choose s}\left[1-q_N(t)\right]^s q_N(t)^{n-s}\,.
\label{S'Nnst}
\ee
The mean value $n[1-q_N(t)]$, which is a sum of single-site terms, is in agreement 
with the exact result~\eref{sNnt}. The binomial distribution gives an extensive 
expression for the variance
\be
\overline{\Delta {s'}_{N,n}^2(t)}=nq_N(t)\left[1-q_N(t)\right]\,,
\label{ds'Nn2t}
\ee
which, due to correlation effects, differs from the exact result in~\eref{dsNn2t}.

In the scaling limit, the rescaled binomial distribution, $n^{1/2}S'_{N,n}(s,t)$, leads to a Gaussian density 
in the scaling variable $\sigma$ defined in~\eref{scals}:
\be
\fP(\sigma,\phi')=\frac{\e^{-\sigma^2/(2\phi')}}{\sqrt{2\pi\phi'}}\,,
\qquad\phi'(w)=\e^{-w}(1-\e^{-w})\,.
\label{fP'}
\ee
The variance $\phi'(w)$ is the rescaled variance $\overline{\Delta {s'}_{N,n}^2(t)}/n$ in the scaling limit.
It differs from $\phi(f,w)$ in~\eref{scalds2dv2} by a term proportional to $f$, of the second order in $e^{-w}$, which is a 
remnant of the correlations. This $f$-dependent correction to the mean-field result becomes less and less important 
as $w$ increases, as shown in figure~\ref{fig-4}(a).

\section{Partial differential equation for the probability density $\fS$}
We start from the master equation~\eref{masteq-1}. Using the scaling variables in~\eref{scal} and~\eref{scals}, the prefactors 
take the following forms:
\be\fl
\frac{N-n+s}{N}\scal 1-f\e^{-w}+\frac{f^{1/2}\sigma}{N^{1/2}}\,,
\qquad \frac{n-s+1}{N}\scal f\e^{-w}-\frac{f^{1/2}\sigma}{N^{1/2}}+\frac{1}{N}\,.
\label{pref-1}
\ee
In the scaling limit $n^{1/2}S_{N,n}(s,t)$ gives the probability density 
$\fS[\sigma(s,t),f,w(t)]$ which depends on $s$ and $t$ through
$\sigma$ and~$w$. A Taylor expansion in the variables $s$ and $t$ 
can be used on the left-hand side of the master equation~\eref{masteq-1} leading to:
\bea
\fl\fS&=\left(1-f\e^{-w}+\frac{f^{1/2}\sigma}{N^{1/2}}\right)\left(\fS-\frac{\partial\fS}{\partial t}
+\frac{1}{2}\frac{\partial^2\fS}{\partial t^2}\right)\nonumber\\
\fl&\ \ \ \ \ \ \ \ +\left(f\e^{-w}-\frac{f^{1/2}\sigma}{N^{1/2}}+\frac{1}{N}\right)
\left(\fS-\frac{\partial\fS}{\partial s}-\frac{\partial\fS}{\partial t}
+\frac{1}{2}\frac{\partial^2\fS}{\partial s^2}+\frac{1}{2}\frac{\partial^2\fS}{\partial t^2}
+\frac{\partial^2\fS}{\partial s\partial t}\right)\,.
\label{pdfS-2}
\eea
 The needed partial derivatives are easily
obtained and read:
\bea
\fl\frac{\partial\fS}{\partial s}&=\frac{1}{N^{1/2}f^{1/2}}\frac{\partial\fS}{\partial \sigma}\,,\qquad
\frac{\partial\fS}{\partial t}=\frac{1}{N}\frac{\partial\fS}{\partial w}
-\frac{f^{1/2}\e^{-w}}{N^{1/2}}\frac{\partial\fS}{\partial \sigma}\,,\qquad
\frac{\partial^2\fS}{\partial s^2}=\frac{1}{Nf}\frac{\partial^2\fS}{\partial \sigma^2}\,,\nonumber\\
\fl\frac{\partial^2\fS}{\partial t^2}&=\frac{f\e^{-2w}}{N}\frac{\partial^2\fS}{\partial \sigma^2}+{\mathrm O}\!\left(N^{-3/2}\right)\,,\qquad
\frac{\partial^2\fS}{\partial s\partial t}=-\frac{\e^{-w}}{N}\frac{\partial^2\fS}{\partial \sigma^2}+{\mathrm O}\!\left(N^{-3/2}\right)\,.
\label{partder-1}
\eea
Note that the second order expansion is sufficient to keep terms up to order $N^{-1}$.
Collecting coefficients of the different powers of $N^{-1/2}$ in~\eref{pdfS-2}, the leading non-vanishing contribution is of order
$N^{-1}$ and gives the partial differential equation~\eref{pdfS-1}.

\section{Partial differential equation for the probability density $\fT$}
With~\eref{scal} and~\eref{scalt}, the prefactors in the master equation~\eref{masteq-3} can be rewritten as:
\be\fl
\frac{N-n+v-1}{N}\scal 1-f(1-x)-\frac{1}{N}\,,\qquad \frac{n-r+1}{N}\scal f(1-x)+\frac{1}{N}\,.
\label{pref-2}
\ee
The scaling limit of $N^{1/2}T_{N,n}(v,t)$ is the probability density $\fT[\tau(t,v),f,x(v)]$ 
depending on $t$ and $v$ through
$\tau$ and $x$. In the continuum limit, a Taylor expansion in~$t$ and~$v$ of the probability 
density on the left-hand side of~\eref{masteq-3} leads to:
\bea
\fl\fT&=\left[1-f(1-x)-\frac{1}{N}\right]\left(\fS-\frac{\partial\fT}{\partial t}
+\frac{1}{2}\frac{\partial^2\fT}{\partial t^2}\right)\nonumber\\
\fl&\ \ \ \ \ \ \ \ +\left[f(1-x)+\frac{1}{N}\right]
\left(\fT-\frac{\partial\fT}{\partial v}-\frac{\partial\fT}{\partial t}
+\frac{1}{2}\frac{\partial^2\fT}{\partial v^2}+\frac{1}{2}\frac{\partial^2\fT}{\partial t^2}
+\frac{\partial^2\fT}{\partial v\partial t}\right)\,.
\label{pdfT-2}
\eea
The partial derivatives have the following expressions:
\bea
\fl\frac{\partial\fT}{\partial t}&=\frac{1}{N^{1/2}}\frac{\partial\fT}{\partial\tau}\,,
\qquad\frac{\partial\fT}{\partial v}=-\frac{1}{N^{1/2}f(1\!-\!x)}\frac{\partial\fT}{\partial\tau}
+\frac{1}{Nf}\frac{\partial\fT}{\partial x}\,,\qquad
\frac{\partial^2\fT}{\partial t^2}=\frac{1}{N}\frac{\partial^2\fT}{\partial \tau^2}\,,\nonumber\\
\fl\frac{\partial^2\fT}{\partial v^2}&=\frac{1}{Nf^2(1\!-\!x)^2}
\frac{\partial^2\fT}{\partial\tau^2}+{\mathrm O}\!\left(N^{-3/2}\right)\,,\quad
\frac{\partial^2\fT}{\partial v\partial t}=-\frac{1}{Nf(1\!-\!x)}
\frac{\partial^2\fT}{\partial\tau^2}+{\mathrm O}\!\left(N^{-3/2}\right)\,.
\label{partder-2}
\eea
Higher derivatives are of higher order in $N^{-1/2}$ and can be ignored in~\eref{pdfT-2}. 
The leading non-vanishing contribution is of order $N^{-1}$ and gives the partial differential equation~\eref{pdfT-1}.

\section{Expression of $\bi{NT_{N,n}(n-u,t)}$}
Inserting the explicit expression of the Stirling numbers of the second kind (equation~(2.9) in~I)
\be
{k\brace m}=\frac{1}{m!}\sum_{j=0}^m(-1)^{m-j}{m\choose j}j^k
\label{stir-1}
\ee 
into~\eref{rStir-2}, one obtains:
\be\fl
{r+l\brace r+m}_r=\frac{1}{m!}\sum_{j=0}^m(-1)^{m-j}{m\choose j}\sum_{k=0}^l{l\choose k}j^k r^{l-k}
=\frac{1}{m!}\sum_{j=0}^m(-1)^{m-j}{m\choose j}(r+j)^l\,.
\label{rStir-5}
\ee
With $T_{N,n}(v,t)$ in~\eref{Tvt-1} one has the correspondence $l=t-1$, $m=v-1=n-u-1$, $r=N-n$, so that:
\bea\fl
NT_{N,n}(n-u,t)&=\frac{n!}{u!N^{t-1}}\frac{1}{(n-u-1)!}\sum_{j=0}^{n-u-1}(-1)^{n-u-j-1}{n-u-1\choose j}(N-n+j)^{t-1}\nonumber\\
\fl&=\frac{n!}{u!(n-u-1)!}\sum_{j=0}^{n-u-1}(-1)^{n-u-j-1}{n-u-1\choose j}\left(1-\frac{n-j}{N}\right)^{t-1}\,.
\label{NTNn-1}
\eea
Finally, changing the sum over $j$ into a sum over $l=n-u-j-1$, equation~\eref{NTNn-2} is obtained.

\end{document}